\documentclass[12pt]{article}
\usepackage{graphicx}
\usepackage{psfig,picinpar,floatflt,amssymb,epsf}
\csname @addtoreset\endcsname{equation}{section}

\def\bseq{\begin{subequation}}  % = 1a 1b
\def\eseq{\end{subequation}}
\def\bsea{\begin{subeqnarray}}  % = 1.1a 1.1b
\def\esea{\end{subeqnarray}}
\newcommand{\bbox}{\lower.2ex\hbox{$\Box$}}

\newcommand{\beq}{\begin{equation}}
\newcommand{\eeq}{\end{equation}}
\newcommand{\bea}{\begin{eqnarray}}
\newcommand{\eea}{\end{eqnarray}}
\newcommand{\ena}{\end{eqnarray}}

\renewcommand{\a}{\alpha}
\renewcommand{\b}{\beta}

\renewcommand{\d}{\delta}
\renewcommand{\th}{\theta}

\newcommand{\g}{\gamma}
\newcommand{\G}{\Gamma}

\newcommand{\e}{\epsilon}

\newcommand{\Db}{\bar{D}}

\newcommand{\Phib}{\bar{\Phi}}

\newcommand{\ad}{{\dot{\alpha}}}

\begin{document}
\begin{titlepage}
\begin{flushright}
Bicocca--FT--05--15 \\
IFUM--838--FT \\
%hep-th/9910197
\end{flushright}
\vspace{.3cm}
\begin{center}
{\Large \bf Two-point correlators in the $\b$--deformed ${\cal N}=4$ SYM at
the next--to--leading order }
\vfill%\vskip 15mm%27.mm

{\large \bf Silvia Penati$^1$,
Alberto Santambrogio$^1$ and
Daniela Zanon$^2$}\\
\vfill%\vskip 7mm%1cm

{\small
$^1$ Dipartimento di Fisica, Universit\`a di Milano--Bicocca
and\\ INFN, Sezione di Milano, Piazza della Scienza 3,
I--20126 Milano, Italy\\
$^2$ Dipartimento di Fisica, Universit\`a di Milano
and\\ INFN, Sezione di Milano, Via Celoria 16,
I--20133 Milano, Italy\\}
\end{center}
\vfill
\begin{center}
{\bf Abstract}
\end{center}
{\small  We compute two--point functions of lowest weight operators at the next--to--leading order 
in the couplings for the $\b$--deformed ${\cal N}=4$ SYM. In particular we focus on the CPO ${\rm 
Tr}(\Phi_1^2)$ and the operator ${\rm Tr}(\Phi_1 \Phi_2)$ not presently listed as BPS. 
We find that for both 
operators no anomalous dimension is generated at this order, then confirming the results recently 
obtained at lowest order in hep-th/0506128. However, in both cases a finite correction to the 
two--point function appears. } \vspace{2mm} \vfill \hrule width 3.cm
\begin{flushleft}
e-mail: silvia.penati@mib.infn.it\\
e-mail: alberto.santambrogio@mi.infn.it\\
e-mail: daniela.zanon@mi.infn.it
\end{flushleft}
\end{titlepage}

Exactly marginal deformations of ${\cal N}=4$ SYM theory preserving ${\cal N}=1$ 
supersymmetry \cite{LS} have recently attracted much attention, in particular in
connection with AdS/CFT correspondence \cite{M}. In fact, the supergravity duals
of the so--called $\beta$-deformations of ${\cal N}=4$ SU(N) SYM theory
have been found by Lunin and 
Maldacena in \cite{LM}. This opens the possibility of making a comparison
between correlation functions of (un)protected operators in the two weak/strong
coupling regimes of the theory, in a way similar to what has been done in 
the last few years for the undeformed ${\cal N}=4$ theory. 

The spectrum of chiral primary operators (CPO) for the deformed theory has been 
found in \cite{BL}. They are classified according to the charge assignement
with respect to three U(1) groups such that $\Phi_i$ is charged under the i-th
group \cite{LM,FG}. For generic deformation parameters the chiral ring is given by 
operators ${\rm Tr}(\Phi_i^J)$, $i=1,2,3$, and 
${\rm Tr}(\Phi_1^J \Phi_2^J \Phi_3^J)$.

In the ${\cal N}=4$ case the operators  ${\rm Tr}(\Phi_i^J)$
and ${\rm Tr}(\Phi_i^{J-1}\Phi_k )$, $k \neq i$ are both CPO and related by 
a SU(3) R--symmetry transformation. Therefore they share 
the same properties: They do not acquire anomalous dimension
and their 2- and 3-point functions are not corrected at the quantum level. 

In the deformed case the two sets of operators belong to two different classes 
and they might undergo a different destiny. It is therefore
compelling to investigate their quantum properties.
This program has been undertaken in a very recent paper \cite{FG} where
the authors have computed 2- and 3-point functions of some operators in the 
deformed superconformal field theory at the leading order in perturbation 
theory (see also \cite{KMSS}). 
In the case of CPO's they find that at this order the corresponding correlators 
have no radiative corrections. This is more than the
expected property of having no anomalous dimension, and it seems to indicate
that these operators are very similar to the CPO's of the undeformed ${\cal N}=4$
theory.\\
However, the most unexpected result of \cite{FG} is that the  
operator Tr($\Phi_1\Phi_2$) has protected 2pt function at 
the lowest order, although it was not recognized as a CPO in the previous literature.
The aim of our paper is to test these unexpected properties at the 
next-to-leading order in perturbation theory, in order to understand if they 
are an accident of the one--loop calculation 
or they signal an actual protection of this operator.

We concentrate on the lowest weight CPO ${\rm Tr}(\Phi_1^2)$ 
and ${\rm Tr}(\Phi_1 \Phi_2)$ and compute their two--point functions at the
next--to--leading order.
We find that the unexpected nonrenormalization of ${\rm Tr}(\Phi_1 \Phi_2)$
persists at the next-to-leading order. This result supports the idea that this
operator should be included in the CPO classification \cite{BL,LM}
\footnote{We thank Juan Maldacena for pointing out the possibility for this
operator to be lacking in the present classification of CPO's for $\b$--deformed
theories.}.

However, we find that the 2pt functions for both operators get a finite correction,
in contradistinction to the CPO's of the undeformed ${\cal N}=4$ case. 

\vskip 30pt

The most convenient setup to perform higher order calculations is the
${\cal N}=1$ superspace (we will use notations and conventions in
\cite{superspace}). In this framework the $\b$--deformed theory is 
described by the following action
\bea
S[J,\bar{J}]
&=&\int d^8z~ {\rm Tr}\left(e^{-gV} \Phib_i e^{gV} \Phi^i\right)+
\frac{1}{2g^2}
\int d^6z~ {\rm Tr} W^\a W_\a\nonumber\\
&&+ih  \int d^6z~ {\rm Tr}( e^{i\pi\b} \Phi_1 \Phi_2 \Phi_3 
- e^{-i\pi\b} \Phi_1 \Phi_3 \Phi_2 )
\nonumber \\
&& + ih^\ast \int d^6\bar{z}~ {\rm Tr} ( e^{i\pi\b} \Phib_1 \Phib_2 \Phib_3
- e^{-i\pi\b} \Phib_1 \Phib_3 \Phib_2 )
\nonumber\\
&&+\int d^6z~ J {\cal O}+\int d^6\bar{z}~ \bar{J}\bar{{\cal O}} \label{actionYM} \eea where $h$ and 
$\b$ are complex couplings. The superfield strength $W_\a= i\Db^2(e^{-gV}D_\a e^{gV})$ is given in 
terms of a real prepotential $V$ and $\Phi_{1,2,3}$ contain the six scalars of the original ${\cal 
N}=4$ SYM theory organized into the ${\bf 3}\times \bf{ \bar 3}$ of $SU(3) \subset SU(4)$. We write 
$V=V^aT_a$, $\Phi_i=\Phi_i^a T_a$ where $T_a$ are $SU(N)$ matrices in the fundamental 
representation. We have added to the classical action source terms for composite chiral operators 
generically denoted by ${\cal O}$ ($J$ ($\bar{J}$) are (anti)chiral sources).

We perform the calculation of two--point correlators by following closely the procedure described in 
\cite{PSZ1,PSZ2,PS}. Here we briefly recall the main steps of the general prescription, while 
referring to those papers for a detailed discussion. 

In Euclidean space we introduce the generating functional \beq W[J,\bar{J}]=\int {\cal D}\Phi~{\cal 
D}\Phib~{\cal D}V~e^{S[J,\bar{J}]} \label{genfunc} \eeq for the $n$--point functions of the 
operator ${\cal O}$ \beq \langle {\cal O}(z_1) \cdots \bar{\cal O}(z_n) \rangle ~=~ \left. 
\frac{\d^n W}{\d J(z_1) \cdots \d \bar{J}(z_n)}\right|_{J=\bar{J}=0} \label{defcorr} \eeq where $z 
\equiv (x,\theta, \bar{\theta})$. The perturbative evaluation of the $n$--point function is 
equivalent to computing    the contributions to $W[J, \bar{J}]$ at order $n$ in the sources. Since 
we are interested in the two-point super-correlator for chiral operators of weight 2 we look for 
quadratic contributions of the form \beq W[J,\bar{J}]\rightarrow \int d^4x_1~d^4x_2~ d^4\theta~ 
J(x_1,\theta,\bar{\theta}) \, \frac{F(g^2,|h|^2, \b,N)}{[(x_1-x_2)^2]^{2 + \g}} \, 
\bar{J}(x_2,\theta,\bar{\theta}) \label{twopoint} \eeq  The $x$-dependence of the result is fixed 
by the conformal invariance of the theory at this order, whereas $F(g^2,|h|^2, \b,N)$ signals 
possible finite quantum corrections and   $\g = \g(g^2,|h|^2, \b,N)$ is the anomalous dimension 
which the operators can acquire at the quantum level.

We work in momentum space, using dimensional regularization and minimal subtraction scheme. In $n$ 
dimensions, with $n=4-2\e$, the naive dimension of the operators is continued to the value 
$2(1-\e)$. The Fourier transform of the power factor in (\ref{twopoint}) is given by \beq 
\frac{1}{(x^2)^{2(1-\e )+\g}} ~=~ 2^{2 -2\g +2\e} \pi^{2-\e} \frac{\G(-\g +\e)} {\G(2(1-\e) +\g)} 
\int \frac{d^n p}{(2\pi)^n} ~\frac{e^{-ipx}}{(p^2)^{-\g+\e}} \label{basicformula2} \eeq By 
performing analytic continuation of $\G(-\g +\e)$ and expanding in powers of $\g$ we obtain (see 
\cite{PS} for details) \beq \frac{1}{(x^2)^{2(1-\e)+\g}} ~\sim~ C \left[ \frac{1}{\e} ~+~ 
\frac{\g}{\e^2} ~+~ \frac{1}{\e} \, O\left( \frac{\g^2}{\e^2}\right) \right] \, \times \int 
\frac{d^n p}{(2\pi)^n} ~ \frac{e^{-ipx}}{(p^2)^{-\g +\e}} \label{basicformula3} \eeq where $C$ is a 
constant. In momentum space and dimensional regularization, the $\frac{1}{\e}$ divergent term 
corresponds to the short distance singularity of the correlation function (\ref{twopoint}) for $x_1 
\sim x_2$, whereas contributions $\frac{1}{\e^2}$ signal the presence of an anomalous dimension. 
Therefore, in performing perturbative calculations in momentum space we will look for all the 
contributions to (\ref{twopoint}) that behave like $\frac{1}{\e}$ and $\frac{1}{\e^2}$, 
disregarding finite contributions (in $x$ space they would give contact terms). Once the divergent 
terms are determined at a given order in the couplings, by anti--Fourier transforming back to the 
configuration space we can reconstruct an $x$--space structure as in (\ref{twopoint}) with a 
nonvanishing contribution to $F(g^2,|h|^2, \b,N)$ and $\g$.

The basic rules of our strategy can then be summarized as follows:
we consider all the two-point diagrams from $W[J,\bar{J}]$ with $J$ and 
$\bar{J}$ on the external legs. We write the corresponding analytic expression
by using Feynman rules as coming from the action (\ref{actionYM}) (after 
gauge--fixing, we work in Feynman gauge where $<VV> = \frac{1}{\Box}$).
Moreover we find convenient to rewrite the chiral superpotential as
\beq
-h( f_{abc} \cos{\pi\b}  + d_{abc} \sin{\pi\b} ) \int d^6z ~\Phi_1^a \Phi_2^b 
\Phi_3^c ~+~ {\rm h.c.}
\label{vertex}
\eeq
where $f_{abc} = -i {\rm Tr}(T_a [T_b ,T_c])$ and 
$d_{abc} =  {\rm Tr}(T_a \{T_b ,T_c\})$ \footnote{We have slightly changed
conventions with respect to our old papers \cite{PSZ1,PSZ2,PS} by 
rescaling the constants $d_{abc}$ by a factor 2.}.
Then we evaluate
all combinatorics factors of a given diagram and compute the colour
structure (we use conventions and identities listed in Appendix A of
\cite{PSZ1}). Then we perform the
superspace $D$-algebra following standard techniques reducing the result 
to a multi-loop
momentum integral. Finally we compute its  $\frac{1}{\e}$ and $\frac{1}{\e^2}$ 
divergent contributions (we refer to Appendix B of \cite{PSZ1} for the 
integrals).

\vskip 20pt

Before entering the explicit calculation of correlators we need investigate
the renormalization properties of the theory described by the action
(\ref{actionYM}). In particular we are interested in the evaluation of the
beta functions up to two loops in order to impose the condition of 
superconformal invariance at the quantum level (vanishing beta 
functions). For the ${\cal N}=4$ SYM theory and in a superspace setup, 
two--loop beta functions were computed in \cite{GRS} where one and two--loop 
diagrams contributing to the propagators and vertices can be found. The
$\b$--deformed theory differs from ${\cal N}=4$ SYM only for the structure
of the chiral vertex, while the propagators and the vector--chiral vertices
are the same. Therefore, all the perturbative contributions from diagrams which
do not contain chiral vertices are the same as for ${\cal N}=4$ and we can
read the results in \cite{GRS}. What we are left with is the calculation of
diagrams with chiral vertices. 

At one--loop order the corrections to the propagators of the fundamental chiral
fields are given in Fig. 1. 
\vskip 18pt
\noindent
%---------- FIGURE TOP ------------
\begin{minipage}{\textwidth}
\begin{center}
\includegraphics[width=0.80\textwidth]{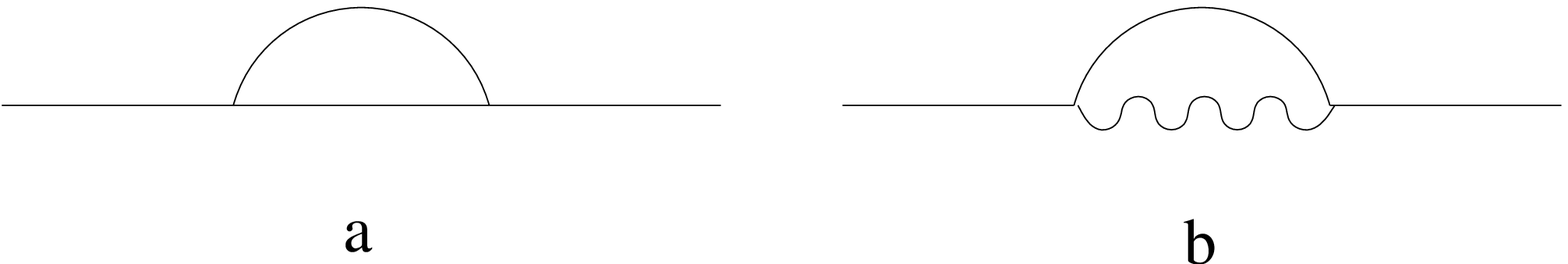}
\end{center}
\begin{center}
{\small{Figure 1: One--loop contributions to the chiral propagator}}
\end{center}
\end{minipage}
%---------- FIGURE END ------------

\vskip 20pt
Computing the first diagram with the chiral vertex (\ref{vertex}) and using 
known results for the second diagram we find the following divergent 
contribution to $\int d^8z {\rm Tr}(\Phi_i \Phib_i)$
\beq
\frac{1}{(4\pi)^2 \e} \, 2N \left[ |h|^2 \left( \cos^2{\pi\b} +
\frac{N^2-4}{N^2} \sin^2{\pi\b} \right) - g^2 \right] 
\label{oneloop}
\eeq
in agreement with \cite{FG}.
This gives rise to an anomalous dimension for the chiral fields proportional 
to the square bracket in (\ref{oneloop}). According to the non--renormalization
theorem for the $N=1$ chiral superpotential \cite{GRS,S} 
the beta function for the chiral 
coupling is proportional to the anomalous dimension of the chiral fields. 
Therefore, one--loop superconformal invariance requires 
\beq
 \left[ |h|^2 \left( \cos^2{\pi\b} +
\frac{N^2-4}{N^2} \sin^2{\pi\b} \right) - g^2 \right] = 0
\label{cond}
\eeq
The same condition insures that also the gauge coupling beta function 
vanishes at this order. This follows from general renormalization properties
of these theories \cite{JJ}, or it can be  proven easily by a direct 
calculation \cite{GRS} of the $V\Phi\Phib$ 3pt function which, under the condition 
(\ref{cond}), turns out to be finite and identical to the one in the ${\cal N}=4$
theory
\bea
&&\frac{g^3}{4}~ N ~k_2~if^{abc} ~\Phib^i_a (q,\theta,\bar{\theta})
\Phi^i_b(-p,\theta,\bar{\theta})
\left( 4D^\a\Db^2D_\a +(p+q)^{\a\ad}[D_\a,\Db_\ad ]\right) 
V_c (p-q,\theta,\bar{\theta}) \nonumber\\
&& ~~~~~~~~~~~~~~~~~~~~~~~~~~~~~~~~  \times \int \frac{d^nk}{k^2(k-p)^2(k-q)^2}
\label{3vertex}
\eea
Moving to two loop order,  first we compute the self--energy corrections 
to the chiral propagators. Exploiting the results for the ${\cal N}=4$ case and
recomputing the contributions of diagrams containing chiral vertices we find 
that, under the condition (\ref{cond}), the result is finite and coincides
with the ${\cal N}=4$ result \cite{GRS}
\bea
&&-2g^4~N^2~k_2 ~\Phib^i_a(p,\theta,\bar{\theta})~ 
\Phi^i_a(-p,\theta,\bar{\theta})~p^2~
\int \frac{d^n q~d^nk}{k^2q^2(k-q)^2(k-p)^2(p-q)^2}\nonumber\\
&&~~~~~~~~~~~=-2g^4~N^2~k_2~ \Phib^i_a(p,\theta,\bar{\theta})~ 
\Phi^i_a(-p,\theta,\bar{\theta})~\frac{1}{(p^2)^{2\e}}
[6\zeta(3)+{\cal O}(\e)]
\label{2prop}
\eea
Therefore the condition (\ref{cond}) is sufficient to guarantee the vanishing
of the anomalous dimension of the fundamental fields at this order. 
Again, according to general renormalization arguments \cite{JJ} this should 
also imply the vanishing
of beta functions, i.e. superconformal invariance at two loops. 

\vskip 20pt
Now armed with the condition (\ref{cond}) we  compute the two--point 
function for the chiral primary operator 
${\rm Tr}(\Phi_1^2)$ up to the next--to--leading 
(two--loop) order. This correlator has been recently computed at one--loop
in \cite{FG} by using a component description of the theory in the WZ gauge. 
The result indicates that at lowest order non only the 
operator is not renormalized as it should be, but also there is no finite
correction to the correlator. In a superspace language this result is very
easy to reproduce. In fact at this order the only diagrams contributing  are
\vskip 18pt
\noindent
%---------- FIGURE TOP ------------
\begin{minipage}{\textwidth}
\begin{center}
\includegraphics[width=0.55\textwidth]{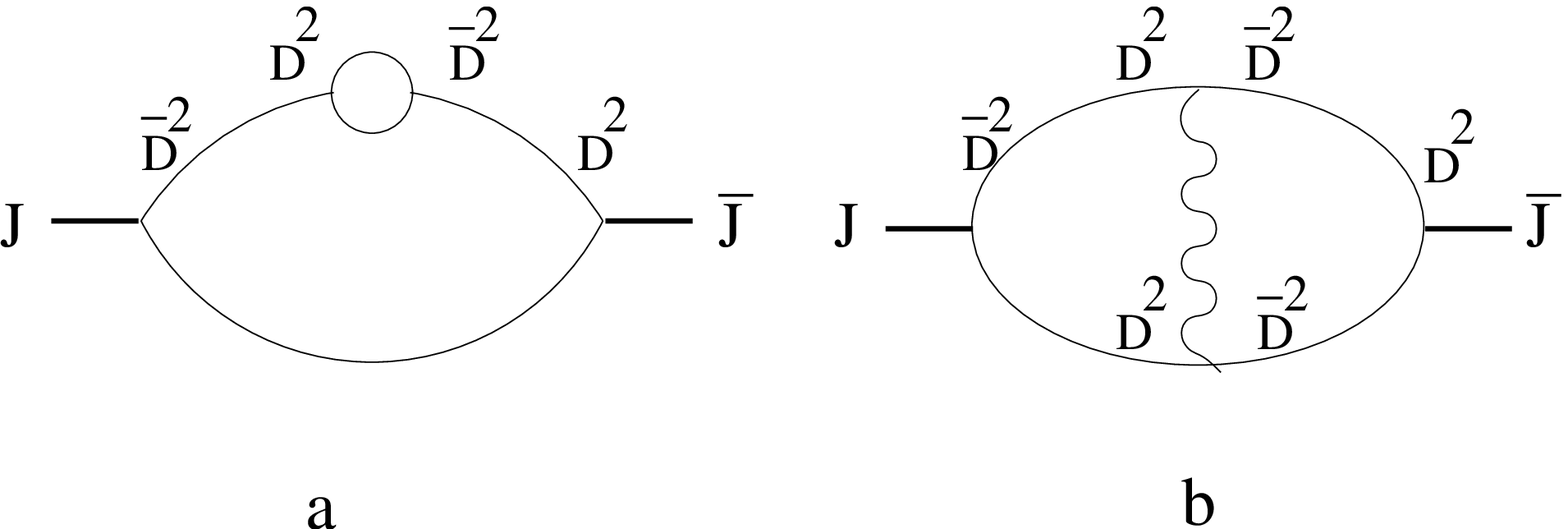}
\end{center}
\begin{center}
{\small{Figure 2:
One--loop contributions to $<{\rm Tr}(\Phi_1)^2 {\rm Tr}(\Phib_1)^2>$}}
\end{center}
\end{minipage}
%---------- FIGURE END ------------

\vskip 20pt
\noindent
where in the first diagram the one--loop correction to the propagator has 
been inserted. Since under the condition (\ref{cond}) this insertion vanishes
the first diagram is trivially zero in the superconformal case. Once D--algebra has been performed
the second
diagram corresponds to a finite (not interesting for us)
momentum integral. Thus we can conclude that
at the lowest order the 2pt function does not get either finite corrections 
or anomalous dimension contributions. 

At two loops the potentially divergent diagrams  are
given in Fig. 3 (we neglect diagrams which were shown to be finite in 
\cite{PSZ1}). 

In diagrams (3a) and (3b) the two--loop chiral self--energy 
(\ref{2prop}) and the gauge--chiral vertex correction (\ref{3vertex}) 
have been inserted, respectively. The results for these diagrams can be 
read from the ${\cal N}=4$ case, as well as the one for diagram (3c) 
(see \cite{PSZ1}). Extracting an overall factor
\beq
16\frac{1}{(4\pi)^6}~N^2(N^2-1)~
\int d^4p~d^4\theta~J(-p,\theta,\bar{\theta})\bar{J}(p,\theta,\bar{\theta})
\label{overall}
\eeq
we have
\bea
&& {\rm Fig.~3a}~\rightarrow~g^4~\zeta(3)\frac{1}{\e}
\nonumber \\
&& {\rm Fig.~3b}~\rightarrow~ -2g^4~\zeta(3)\frac{1}{\e}
\nonumber \\
&& {\rm Fig.~3c}~\rightarrow~\frac{1}{2}g^4~\zeta(3)\frac{1}{\e}
\label{3equal}
\eea

\vskip 18pt
\noindent
%---------- FIGURE TOP ------------
\begin{minipage}{\textwidth}
\begin{center}
\includegraphics[width=0.50\textwidth]{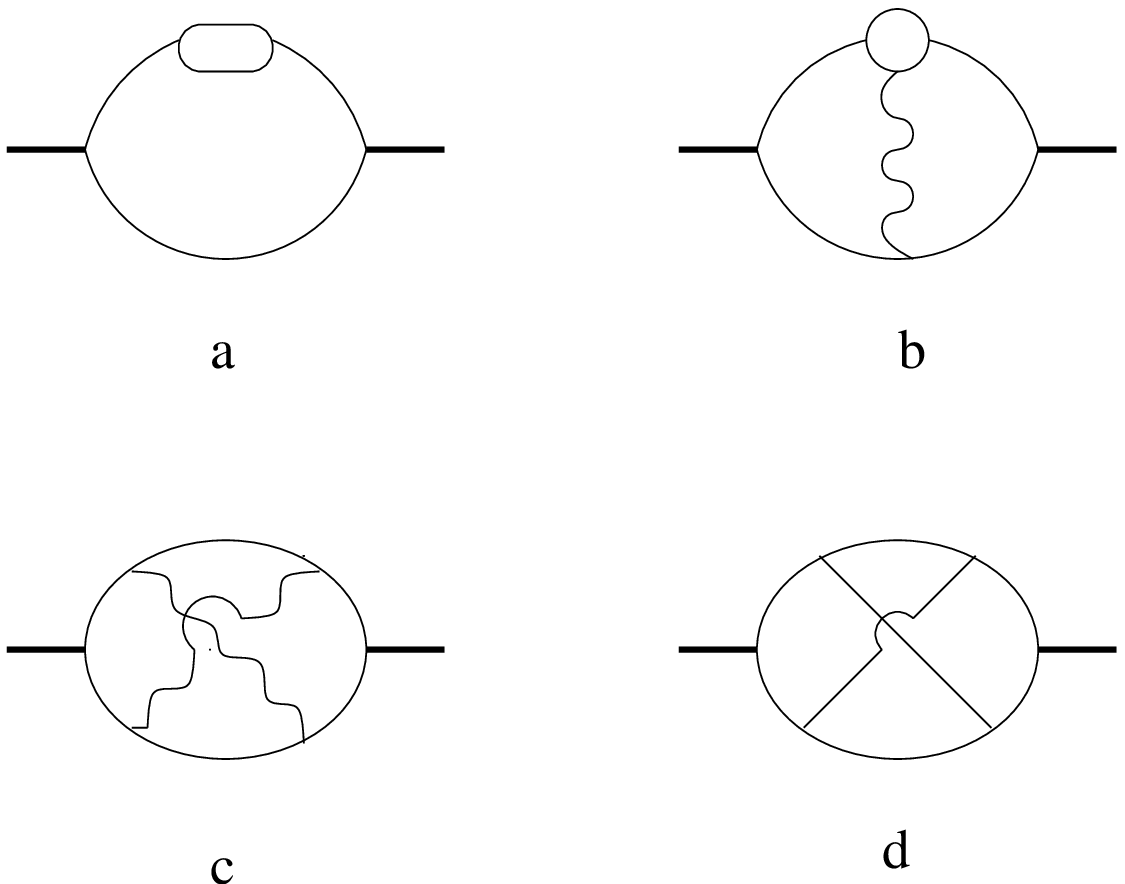}
\end{center}
\begin{center}
{\small{Figure 3: Two--loop
contributions to $<{\rm Tr}(\Phi_1)^2 {\rm Tr}(\Phib_1)^2>$}}
\end{center}
\end{minipage}
%---------- FIGURE END ------------

\vskip 20pt

Diagram (3d) needs to be computed since it contains the new vertex 
(\ref{vertex}). What changes is the color factor, while the D--algebra and the
momentum integral are identical to the ${\cal N}=4$ case. Again, extracting
the overall factor (\ref{overall}) and using (\ref{cond}) we obtain
\bea
&& \frac12 |h|^4~\left[ \cos^4{\pi\b} - 6 \frac{N^2-4}{N^2}\cos^2{\pi\b}\sin^2{\pi\b}
+ \frac{(N^2-4)(N^2-12)}{N^4}\sin^4{\pi\b} \right] \zeta(3) \frac{1}{\e} 
\nonumber \\
&=& \frac{1}{2}\left[ g^4 + 8 |h|^4 \frac{N^2-4}{N^2}\sin^2{\pi\b}
\left( \frac{N^2-1}{N^2}\sin^2{\pi\b} - 1 \right)  \right]~\zeta(3)\frac{1}{\e}
\label{4equal}
\eea
where the condition (\ref{cond}) has been used once again.
Summing up all the contributions we immediately see that the $\frac{1}{\e}$ poles
proportional to $g^4$ cancel in agreement with the ${\cal N}=4$ case, whereas
a new non vanishing contribution proportional to $|h|^4$ appears. This term is
proportional to $\sin^2{\pi\b}$ and it vanishes in the undeformed limit 
($\b \to 0$, $|h|^2 \to g^2$). When we transform the result back to 
configuration space we end up with a nontrivial correction to the 2pt function
of order $|h|^4$
\beq
\langle {\rm Tr}(\Phi_1^2)(z_1) {\rm Tr}(\Phib_1^2)(z_2) \rangle_{\rm 2-loops} 
~\sim~
\frac{\d^{(4)}(\th_1 - \th_2)}{[(x_1 - x_2)^2]^2} ~
|h|^4 \frac{N^2-4}{N^2}\sin^2{\pi\b}
\left( \frac{N^2-1}{N^2}\sin^2{\pi\b} - 1 \right) 
\label{11}
\eeq 
We note that this correction survives even in the large $N$ limit. 

\vskip 20pt
Now we consider the operator ${\rm Tr} (\Phi_1 \Phi_2)$ which in principle
might renormalize. In \cite{FG} it has been shown that at the lowest
order in perturbation theory this operator does not acquire anomalous 
dimension and its 2pt function is not corrected. 
This result can be easily reproduced in our language: The one--loop
diagrams contributing to the two--point function for this operator are still 
given in Fig. 2. Since the first diagram is zero in the superconformal case 
while the second diagram is always finite, we do not get either corrections 
to the correlators or anomalous dimension contributions. 

In order to investigate whether the absence of anomalous dimension is an
accident of the one--loop calculation we compute the next--to--leading 
contributions to the 2pt function. The diagrams contributing to this 
correlator are still the ones given in Fig. 3 where now we have $\Phi_1$ and
$\Phi_2$ fields coming out from the source vertices, plus the extra diagram
in Fig. 4 where we have indicated explicitly the two possible contractions
of $\Phi_1, \Phi_2$ and $\Phib_1, \Phib_2$.    
 
\vskip 18pt
\noindent
%---------- FIGURE TOP ------------
\begin{minipage}{\textwidth}
\begin{center}
\includegraphics[width=0.60\textwidth]{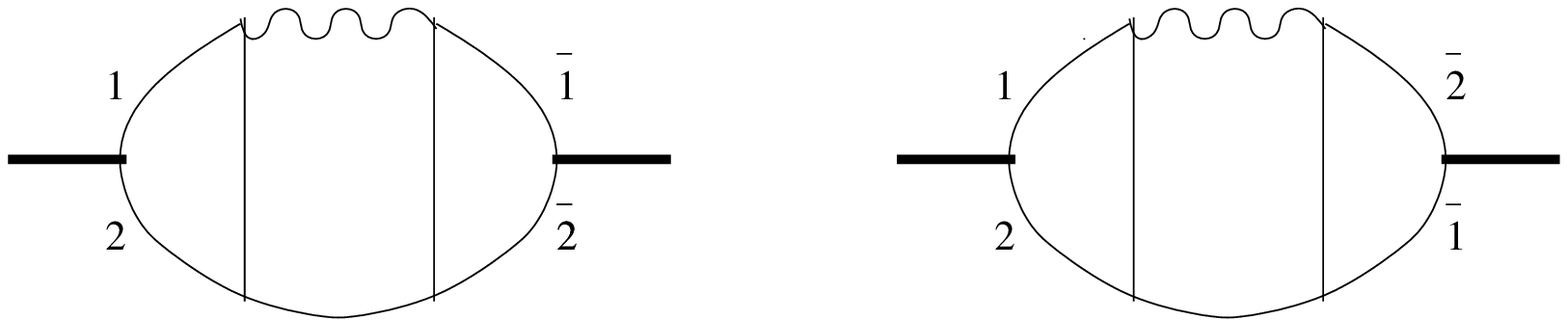}
\end{center}
\begin{center}
{\small{Figure 4: Additional two--loop
diagram for $<{\rm Tr}(\Phi_1 \Phi_2) {\rm Tr}(\Phib_1 \Phib_2)>$}}
\end{center}
\end{minipage}
%---------- FIGURE END ------------

\vskip 20pt 

In computing the contributions from Fig. 3 one easily realizes that what
changes with respect to the previous calculation is the color factor of diagram
(3d), while D--algebra and momentum integrals remain the same. Therefore, 
as it immediately appears from eqs. (\ref{3equal}, \ref{4equal}) we will 
still get at most $1/\e$ divergences, so no contributions to the anomalous
dimension arise from the diagrams in Fig. 3. The only potential source of 
$1/\e^2$ contributions is the diagram in Fig. 4 since, after completion of the
D--algebra, it gives rise to momentum loop integrals with self--energy 
subdivergences. However, there is a complete cancellation between the 
two contributions drawn there due to a sign change in the color structure
produced by the exchange of $\Phib_1$ with $\Phib_2$. We then conclude that
the diagram in Fig. 4 does not contribute and even at this order the 
operator ${\rm Tr}( \Phi_1 \Phi_2)$ does not acquire an anomalous dimension. 
Our result confirms the lowest order \cite{FG} protection of this operator,
then giving further support to its BPS nature. 

Contributions $1/\e$ from diagrams in Figs. (3a, 3b, 3c) are still given
in eqs. (\ref{3equal}). From diagram (3d) we obtain
\bea
&& \frac12 |h|^4~\left[ \cos^4{\pi\b} + 2 \frac{N^2 -4}{N^2} 
\cos^2{\pi\b}\sin^2{\pi\b} + \frac{(N^2-4)(N^2-12)}{N^4} \sin^4{\pi\b} \right]
\zeta(3) \frac{1}{\e}
\nonumber \\
&=& \frac12 \left[ g^4 - 8 |h|^4 \frac{N^2-4}{N^4} \sin^4{\pi\b}  \right]
\zeta(3) \frac{1}{\e}
\eea
where we have used (\ref{cond}). 
Again, we see that the $g^4$ contributions cancel, consistently with 
the ${\cal N}=4$ case. We are left with a contribution proportional to
$|h|^4$ which gives a 
finite correction to the two--point function of the operator
\beq
\langle {\rm Tr}(\Phi_1\Phi_2)(z_1) {\rm Tr}(\Phib_1\Phib_2)(z_2) 
\rangle_{\rm 2-loops} ~\sim~
\frac{\d^{(4)}(\th_1 - \th_2)}{[(x_1 - x_2)^2]^2} ~
|h|^4 \frac{N^2-4}{N^4} \sin^4{\pi\b}
\label{12}
\eeq 
This contribution vanishes in the undeformed limit and it is 
subleading in the large $N$ limit. 

\vskip 10pt

The results presented in this letter are part of a systematic analysis of correlators
in the deformed ${\cal N}=4$ SYM theory \cite{MPSZ}.

\medskip

\section*{Acknowledgements}
\noindent We wish to thank Dan Freedman for discussions and
helpful suggestions and Juan Maldacena for enlightening comments.\\
\noindent This work has been supported in
part by INFN, PRIN prot. 2003023852\_008 and the European Commission RTN 
program MRTN--CT--2004--005104.

\end{document}